\documentstyle[12pt]{article}
\def\be{\begin{equation}}
\def\ee{\end{equation}}
\def\ba{\begin{eqnarray}}
\def\ea{\end{eqnarray}}
\def\ga{\mathrel{\raise.3ex\hbox{$>$\kern-.75em\lower1ex\hbox{$\sim$}}}}
\def\la{\mathrel{\raise.3ex\hbox{$<$\kern-.75em\lower1ex\hbox{$\sim$}}}}

\newcommand{\bi}[1]{\bibitem{#1}}
\newcommand{\fr}[2]{\frac{#1}{#2}}

\begin{document}

\begin{titlepage}
\samepage{
\setcounter{page}{0}
\rightline{TPI--MINN--00/11}
\rightline{UMN--TH--1845--00}
\rightline{March 2000}
\vfill
\begin{center}
 {\Large \bf Direct and indirect limits on the electro-magnetic 
form factors of WIMPs.}
\vfill
\vspace{.25in}
 {\large Maxim Pospelov $\,$and$\,$ Tonnis ter Veldhuis\\}
\vspace{.25in}
 {\it  Department of Physics,
              University of Minnesota, Minneapolis, MN  55455, USA\\}
\end{center}
\vfill
\begin{abstract}
  {\rm

We use the results of direct and indirect searches for
weakly interacting massive particles (WIMPs) to obtain
bounds on various electro-magnetic form factors of 
WIMPs. The limits on the recoil
signal in underground dark matter detection experiments, 
with standard assumptions about the density and r.m.s. velocity of 
WIMPs in the halo, can be translated into
the following model-independent bounds on the magnetic dipole, 
electric dipole, quadrupole and
anapole moments, the charge radius, and the polarizability of WIMPs: 
$\mu_D/\mu_n < 1.4\: 10^{-4}$, 
$d < 5\: 10^{-21}{\rm e}\,{\rm cm}$,  $Q< \:6\: 10^{-7}\, {\rm e}\, {\rm fm}^2$,
$a/\mu_n<3\:10^{-2}\, {\rm fm}$, $r_D^2 < \:9\: 10^{-7}\,{\rm fm}^2$, and 
$\chi < 5\:10^{-8}\,{\rm fm}^3$ for a WIMP mass of
$100$GeV. The limits on fluxes of highly energetic neutrinos produced in the 
annihilation of WIMPs in the center of the earth and the sun lead to somewhat
stronger, but more model dependent bounds.
}
\end{abstract}
\vfill
\smallskip}
\end{titlepage}

\setcounter{footnote}{0}


\setcounter{footnote}{0}

\section{Introduction}

Various observational facts, notably the dynamics of spiral galaxies, clusters
and superclusters, and their theoretical explanations
in modern cosmology and astrophysics can be reconciled 
only by postulating the existence of the 
non-luminous matter at amounts exceeding the total 
matter density provided by baryons. Thus, if gravity does not undergo some
drastic changes at distances larger than few kpc
(a highly improbable option from various points of view),
the explanation of the rotational curves of spiral galaxies
requires the existence of halos, composed of dark matter objects. 

One of the most intriguing options from the
point of view of particle physics is the possibility that
dark matter is composed of massive particles of yet 
unknown identity.
On the one hand these (presumably) weakly interacting massive particles (WIMPs) 
are the subject of intense experimental searches at numerous
underground detectors \cite{Cad}. On the other hand, some of the 
extensions of the Standard Model naturally incorporate such 
stable particles. 
For example, in supersymmetric extensions of SM the 
lightest supersymmetric particle can be stable and neutral, thus 
providing a viable dark matter candidate. Extensive theoretical 
studies of stable SUSY particles, in particular neutralinos, 
have been performed over the last two decades \cite{GJK} for wide ranging values of 
the supersymmetric masses and couplings. We believe, however, that unless 
supersymmetry is experimentally verified  and the stability of 
the lightest superpartner is established in collider
experiments, other possibilities for WIMPs 
should not be discarded.

In this letter we adopt a completely different, yet fully justified,  
phenomenological approach to the physics of WIMPs. 
We do not make any particular  
assumptions about the origin and possible identity of these particles, 
apart from the assumption that their masses are larger than a few GeV. 
We then assess what can be inferred about the properties of WIMPs
directly from  experimental limits implied by various searches.
An additional incentive to address WIMP physics in a model independent
way stems from recently discovered problems with
subgalactic structure formation within the non-interacting cold dark matter 
scenario \cite{problem}. A ``generalized'' WIMP with sufficient 
self-interaction may avoid these problems \cite{selfint}.

Among important low-energy properties of WIMPs are their 
electromagnetic form factors. It has been known for a long
time that the possibility of charged WIMPs is strongly disfavored 
\cite{charge} and stringent limits exist
in the case of a fractional charge \cite{CD}. Here we assume that 
WIMPs are neutral and discuss other electro-magnetic form
factors such as magnetic and electric dipole moments, anapole moment and
polarizability. We put strong constraints on 
the possible 
electromagnetic form factors
of WIMPs using the result of direct
experimental searches aimed at the detection of the 
WIMP-nucleus recoil signal. These constraints depend on the number
density and average velocities of WIMPs in the halo. 
The limits on fluxes
of energetic neutrinos, originating from the annihilation of 
WIMPs in the center of the Sun and the Earth, can also be used
to obtain indirect constraints on the form factors. These constraints 
are less certain, as they require additional assumptions about annihilation 
rates and neutrino branching ratios. 

This paper is organized as follows. The calculations of the WIMP-nuclei 
cross sections are performed in the next section. In section 3 we present direct 
limits on the form factors, using the results of various underground
experiments. In section 4 we obtain indirect limits on the form factors,
using limits on neutrino fluxes. Our conclusions are reserved for
section 5.

\section{Electromagnetic form factors of WIMPs}

The interaction of any compact object and a slowly varying 
electro-magnetic field can be parametrized via a set of 
electro-magnetic form factors. In our case, the object is a WIMP with spin $S$,
and its interactions, linear in external fields $\bf{B}$ and $\bf{E}$, can be 
written as
\ba
H = -\mu {\bf B}\cdot \fr{{\bf S}}{S} 
 -d {\bf E}\cdot \fr{{\bf S}}{ S} - a{\bf j}\cdot{{\bf S}\over S} \\
-{1\over 4 S(2S-1)}[S_iS_j+S_jS_i-{2\over 3}\delta_{ij}S(S+1)]
\left(Q{\partial\over \partial x_i} E_j+
M{\partial\over \partial x_i}B_j\right)+....\nonumber
\ea
Here $\mu$ and $d$ are magnetic and electric dipole moments, and
$M$ and $Q$ are magnetic and electric quadrupole moments.
The anapole moment $a$ is the form factor which 
describes the  contact interaction with the external
current density ${\bf j}$ \cite{ZV}. 

If the spin of the WIMP is zero, these moments do not exist and 
the interaction with electromagnetic field is given by the charge
radius $r_D$ of the WIMP and the  polarizabilities
\be
H=
-\frac{1}{6} e r_D^2 \frac{\partial}{\partial x_i} E_i
- \frac{1}{2}\chi_E E^2 - \frac{1}{2} \chi_B B^2 - \chi_{EB} 
{\bf E}\cdot {\bf B}+... 
\ee

If time reversal is a good symmetry, then $d$, $M$ and $ \chi_{EB}$ necessarily vanish.
Conserved parity also forbids the existence of $a$. Here we do not 
assume that these symmetries are preserved a priori, and we allow all of the 
form factors 
to exist.

In what follows we calculate the cross section of the elastic scattering of 
a WIMP off a nucleus due to the existence of 
the electromagnetic form factors. 
We start with the scattering due to the magnetic 
moment which is arguably the most interesting case. 

The calculation can be done in the non-relativistic limit, 
as $v/c \sim 10^{-3}$ on average for
WIMPs in the halo. We assume that the possible size of the 
form factors is of the order of, or smaller than, the respective 
characteristic values for nuclei,
$\mu<\mu_n$, $Q < e\: {\rm fm}^3$, and so on. It can then easily be
checked that this range 
of velocities also ensures the applicability of the Born approximation.
In our units, $\hbar =1$, $c=1$, $e^2=\alpha=1/137$ and $\mu_n=e/2m_p$.

The cross-section for the scattering of a Wimp 
through the interaction of its magnetic moment with the
magnetic moment of the nucleus is
\be
\sigma = \frac{32 \pi}{3} M_R^2(N,D) \mu_D^2 \mu_N^2 
\frac{\left( S+1 \right) \left( I+1\right)}{3 S I},
\label{magmom}
\ee
where $M_R(N,D)$ is the reduced mass of the WIMP-nucleus system,
$\mu_D$ and $\mu_N$ are the magnetic moments of the WIMP
and the nucleus, respectively, and $I$ is the spin of the
nucleus. The cross-section for the interaction between
the electric dipole moment of the WIMP and the charge of
the nucleus is
\be
\sigma = 8 \pi Z^2 \left( \frac{d}{e}\right)^2 
\left(\frac{\alpha}{v}\right)^2 \frac{S+1}{3S}
\ln \frac{q_{max}}{q_{min}}.
\label{eldip}
\ee
It is logarithmically divergent at large distances. In practice,
$q_{max}$ can be chosen to be of the order of $M_R(N,D)v$ and 
$q_{min}$ is determined either by the inverse of the 
impact parameters at which the nuclear charge is screened by the 
electrons or by the lowest momentum transfer that can
be detected in the experiment, whichever is larger. 
The cross-section for the interaction between the nuclear 
electric current and the anapole of the WIMP is given by
\be
\sigma = \frac{1}{2 \pi} Z^2 \alpha^2 \frac{M_R^2(N,D)}{4 m_p^2}
\left(\frac{a}{\mu_n}\right)^2 v^2 \frac{S+1}{3S},
\label{anmom}
\ee
where $\mu_n$ is the nuclear magneton, and $m_p$ is the proton mass. 
The contribution of the nuclear spin current is negligibly small.
The cross-section for 
scattering through the interaction between the electric quadrupole 
moment of the WIMP and the electric charge of the nucleus takes the
form
\be
\sigma=\frac{8 \pi}{9} M_R^2(N,D) Z^2\alpha^2 \left(\frac{Q}{e}\right)^2  
\frac{\left(S+1\right)\left(2S+3\right)} {10 S \left(2S-1\right)}.
\label{qmom}
\ee
In Eqs. (\ref{eldip})-(\ref{qmom}) the normalization of the 
WIMP spin dependence is such that it is equal to 1 for the lowest spin
for which these moments may exist, $(S+1)/3S=1$ for $S=1/2$ and 
$(S+1)(2S+3)/10S(2S-1)=1$ for $S=1$.
The cross-section for scattering of a WIMP because of its
charge radius in the electric field of the nucleus is given by
\begin{equation}
\sigma=\frac{4 \pi}{9} M_R^2(N,D) Z^2 \alpha^2 r_D^4.
\end{equation}
Finally, the cross-section for the scattering of a WIMP through
its polarizability in the electric field of the nucleus, is
\be
\sigma \simeq \frac{144 \pi}{25} M_R^2(N,D) Z^4\alpha^2 
\frac{\chi_E^2}{r_0^2}.
\label{polariz}
\ee
This cross-section diverges for a point-like nucleus.
We therefore took the charge distribution of the nucleus
to be homogeneous in a sphere with radius $r_0 \sim 1.2\: {\rm fm}\:
\sqrt[3]{A}$.

Scattering due to the magnetic quadrupole moments and magnetic and
mixed  polarizabilities, $\chi_{EB}$, was also considered. These
effects turned out to be additionally suppressed and do not lead to 
interesting limits. We therefore do not quote them here. 

\section{Direct constraints}

We illustrate the bounds on the electro-magnetic form factors of WIMPs
which are implied by the limit on the recoil signal with 
results of the NaI detector of the DAMA experiment \cite{Dama}, 
and Ge detector of the CDMS collaboration \cite{CDMS}. 
Other experiments aimed at the direct detection of dark matter 
can be used as well \cite{UKdata,HDM,Zarag} 
to get limits which would typically be a few times weaker 
(see also Ref. \cite{Cad} for more complete listing of experiments). 
Traditionally, the  results of the direct searches
are presented in the form of a normalized bound on
the \lq\lq WIMP--proton\rq\rq cross-section in the 
spin dependent case, and a normalized \lq\lq WIMP--nucleon\rq\rq 
cross-section in the spin independent case. 
Such normalized bounds are useful to compare results for different
types of nuclei, to compare results from different experiments, and to compare
theoretical predictions with experimental data, but for
our purpose of determining bounds on the WIMP electro-magnetic form
factors, it is necessary to reverse this normalization. The bounds, quoted
in \cite{Dama,UKdata} are, in some sense, specifically formulated for 
neutralino-nucleon scattering, which necessarily implies a nuclear theory
input when switching from nucleus-neutralino to nucleon-neutralino description.
It is clear that in the case of the scattering due to the WIMP
electro-magnetic form factors nuclear theory is not involved, and much of the 
nuclear physics uncertainty is not there. 
Thus, in order to obtain  bounds on the form factors, we recover
the nucleus-WIMP cross section
from the nucleon-WIMP 
by the use of the equation \cite{Dama}
\be
\sigma = \frac{M_R^2\left(N,D\right)}{M_R^2\left(p,D\right)} 
\frac{I(N)}{I(p,n)} \sigma_{p,n} .
\ee
Here $p$ and $n$ stand for proton and neutron, respectively, and the
ratio of the reduced masses takes into account a kinematic factor. In the
spin dependent case, $I(N)/I(p)$ is a spin factor, which takes the values
$I(N)/I(p)=0.009$ for Iodine, and $I(N)/I(p)=0.055$ for Sodium 
\cite{UKdata,EF}.
In the spin independent case, $I(N)/I(n)=(A-Z)^2$ for Ref. \cite{UKdata} and
 $I(N)/I(n)=A^2$ for Ref. \cite{Dama}. For simplicity, we set all
nuclear form factors equal to one, which is a better approximation 
for Sodium than for Iodine.\footnote{For a more rigorous treatment 
electric and magnetic  form factors of nuclei should be included, for
many of which the experimental results are available.} 
A more accurate analysis could be done by the experimental
collaborations using their original data. 
The ensuing bounds on the WIMP electro-magnetic form factors are listed in
Table(\ref{table}). 

\begin{table}
\begin{center}
\begin{tabular}{|c||c|c|}
\hline
      & Na I & Ge \\
\hline \hline
 & & \\
$\left| \mu_D/\mu_n \right|  \sqrt{\left(S+1\right)/3S}$ & $1.4\: 10^{-4}$  & 
  $--$
  \\    & & \\
$\left| d/e \right| \sqrt{\left(S+1\right)/3S}$  & $8\: 10^{-21} {\rm cm}$
   &  $5\: 10^{-21} {\rm cm}$ \\ 
   & & \\
$\left| Q/e \right| \sqrt{\left( S+1 \right) \left(2S+3 \right)/
10S\left( 2S-1\right)}$  
 &  $1\: 10^{-6}{\rm fm}^2$ & $6\: 10^{-7} {\rm fm}^2$ \\ 
 & & \\
$\left| a/\mu_n \right| \sqrt{\left(S+1\right)/3S}$  & $4 \: 10^{-2}{\rm fm}$ &
   $3 \: 10^{-2} {\rm fm}$ \\    & & \\
 $r_D^2$ & $1.4 \: 10^{-6} {\rm fm^2}$ & $ 9\: 10^{-7} {\rm fm^2}$\\ & & \\
$\left|\chi_E \right|$ & $5\: 10^{-8}{\rm fm}^3$ & $7\: 10^{-8}{\rm fm}^3$ \\
  & & \\ \hline 
\end{tabular}
\end{center}
\caption{Bounds on the electro-magnetic form factors for WIMPs with a
mass of $100\: {\rm GeV}$, based on data from the DAMA collaboration 
presented in Ref. \cite{Dama} and from the CDMS collaboration presented
in Ref. \cite{CDMS}.}
\label{table}
\end{table}

The bounds presented in this table are obtained with the standard assumptions 
that the mass density of WIMPs in our halo is 0.3 Gev/cm$^3$ and that their 
velocities are of the order of $200 $ km/sec. The data obtained
with germanium detectors are sensitive to all electro-magnetic form factors
discussed here except the magnetic dipole moment, as the most abundant germanium
isotope has no spin. Let us assume that  
the recently reported annual modulation of the DAMA signal, consistent with 
$m_D\sim 50$ GeV  \cite{DAMAm}, and the absence of the signal, 
resulting in strong limits on spin-independent cross section, 
quoted by CDMS \cite{CDMS}, 
are both correct. Then the presence of a WIMP 
magnetic moment at the level of few$\,\times\, 10^{-5}$ -- $1\times 10^{-4}$ 
compared to the size of the 
nuclear magneton can reconcile the two experimental results. 

It is instructive to consider how these bounds are relaxed in the scaling 
regime, when the WIMP mass is larger than nuclear masses and their abundance 
is kept as a free parameter. To this end
we introduce the parameter $\delta_D$, characterizing the fraction 
of WIMPs of species $D$ in the halo dark matter density ($\delta_D=1$ in Table 1).
The counting rate in a detector is proportional to the number density 
of the WIMP particles and therefore should scale with $M_D$ and $\delta_D$ 
according to the following formula: 
\be
R_{N,D}(\delta_D, M_D) = R_N^0 \delta_D \left(\fr{M_D^0}{M_D}\right),
\ee
where $M_D^0$ is the lowest mass of WIMPs at which the scaling sets in 
and  $R_N^0$ is the counting rate, corresponding to $M_D = M_D^0$ and 
$\delta_D=1$. It is easy to see  from the
figures in Refs.\cite{Dama,CDMS} that $M_D=100$ GeV is close to 
the mass where the scaling behaviour begins \cite{Dama,CDMS}, and therefore we can
simply rescale the bounds given in Table 1 for an arbitrary $\delta_D$ and 
arbitrary and heavy $M_D$, by multiplying them by 
$\sqrt{\delta_DM_D/100{\rm GeV}}$. 

\section{Indirect constraints}

The elastic scattering of WIMPs in the solar (and earth) media leads to  
gravitational trapping. Due to the scattering, WIMPs dissipate energy and
eventually they accumulate in the 
center of the sun (and earth). Their density grows until the accumulation
rate is counter-balanced by the pair annihilation rate, and an equilibrium is
reached. 
In the process of annihilation, among other SM products, 
energetic neutrinos can be produced. The interaction of these
neutrinos in the earth in turn leads to the production of 
energetic
muons, which can be detected with neutrino telescopes.

Thus limits on the flux of energetic neutrinos can be used to 
set bounds on the electro-magnetic form factors of WIMPs. For WIMPS interacting
through their electro-magnetic form factors, the trapping rate
in the center of the sun and earth is determined from the
elastic WIMP-nucleus cross sections that we calculated in
Eqs.(\ref{magmom})-(\ref{polariz}). 

We use 
the results of Ref. \cite{FOP}, which follows the original treatment in 
Refs. \cite{sos},
and estimate the neutrino flux at the surface of the earth due to
WIMP annihilation in the sun as
\be
\phi_{\nu\odot}\simeq (560 cm^{-2}s^{-1})
N_{eff}\delta_D\sigma_{p,36}\fr{\mbox{GeV}^2}{M_D^2}.
\label{flux}
\ee
In this formula $N_{eff}$ is the average number of neutrinos per
annihilation event and $\sigma_{p,36}$ is the WIMP-proton 
elastic cross section in
units of $10^{-36}$ cm$^2$. It is assumed that the equilibrium number of
WIMPs in the sun has been reached,
so that the annihilation and capture rates are equal. 

The experimental limit on the flux of energetic upward
muons (for example obtained by the Kamioka, MACRO and Baksan experiments, 
\cite{Kam,Mac,Bak}),
is $\phi_\mu \la 1.4\times  10^{-14}$ cm$^{-2}$s$^{-1}$, and the probability 
that a neutrino directed towards the detector produces a muon at the
detector is $P(100 {\rm GeV})\sim 10^{-7}$ 
\cite{GS}.
Using Eq.(\ref{flux}), we deduce the following indirect limits on the 
electro-magnetic form 
factors of 
WIMPs with $m_D=100$ GeV,
\begin{eqnarray}
\left| \mu_D/\mu_n \right| \sqrt{\left(S+1\right)/3S} &\la&5~10^{-6} 
\nonumber \\
\left| d/e \right|  \sqrt{\left(S+1\right)/3S} &\la& 2~10^{-22}~ {\rm cm}
\nonumber \\
\left| Q/e \right| \sqrt{\left( S+1 \right) \left(2S+3 \right)
/10S\left( 2S-1\right)}  &\la& 10^{-8} ~{\rm fm}^2 \\
\left| a/\mu_n \right| \sqrt{\left(S+1\right)/3S}  &\la& 
10^{-2}~ {\rm fm} \nonumber \\
r_D^2 & \la & 1.4~10^{-8}~ {\rm fm}^2 \nonumber\\
\left|\chi \right|  &\la& 5.6~ 10^{-7}~ {\rm fm}^3 , \nonumber
\end{eqnarray}
where $N_{eff}$ is taken to be 0.1 and $\delta_D=1$. Note that for higher WIMP masses 
the absorption of neutrinos in the core of the sun becomes important \cite{RS}. 

If the WIMP-nucleus elastic scattering cross-sections in the earth
are comparable to the WIMP-proton cross-section, than
the neutrino signal from the center of the earth is approximately
four orders of magnitude weaker than the signal from the sun \cite{FOP}.
However, the limit on $\chi_E$ from the neutrino flux from the center of
the earth is almost one order of magnitude better
than the limit derived from the neutrino flux from the sun, as the corresponding 
WIMP-nucleus elastic scattering cross-sections are 
enhanced by a large factor, $Z^4\sim 5~10^{5}$, for heavy (Fe) nuclei (see Eq.({\ref{polariz}))
comprising the core of the earth.

Although indirect bounds are definitely stronger, they do require  
a number of additional assumptions. In particular, these bounds 
require that annihilation actually occurs, which in practice may not be true. 
For example, if a WIMP caries a conserved quantum number, 
and there are no ``anti-WIMPs'' - such as heavy Dirac neutrinos in the 
absence of antineutrinos - the annihilation is forbidden.

\section{Discussion}

We have shown that limits on recoil signals from  underground 
detectors can be translated into  bounds on the electro-magnetic form 
factors of WIMPs. Similar bounds can be
inferred from the limits on energetic neutrino fluxes 
from the center of the earth and the sun, generated by annihilating WIMPs.
Our analysis shows that all electromagnetic form factors have to be 
very small when expressed in characteristic nuclear units.
In particular, the magnetic moment of a WIMP has to be five orders of 
magnitude smaller than the magnetic moment of a proton.

Unfortunately, all quoted bounds are trivially satisfied for the 
case of neutralino dark matter. Indeed, because it is a spin-1/2 particle, a
neutralino cannot possess a quadrupole moment. In addition,
because it is a Majorana fermion, a neutralino cannot have  
a magnetic or electric dipole moment either ($\bar\chi \sigma_{\mu\nu} 
\chi\equiv 0$). Finally, an anapole moment and a polarizability can be
generated only radiatively. This results only in a small correction to the 
dominant neutralino-nucleus elastic cross section induced by squark
exchange \cite{GJK}. 

A magnetic moment can arise naturally if WIMPs are Dirac fermions. 
If we assume for a moment that the WIMP is a heavy Dirac neutrino, 
a large magnetic 
moment can be induced due to virtual $W$-charged lepton exchange. 
The value of the induced magnetic moment grows quadratically 
with the size of the Yukawa coupling  until the point where the
Yukawa interaction becomes essentially strong. In latter case 
a natural estimate for the size of the magnetic moment 
is $e/m_\nu$, with $m_\nu$ presumably of the order of 1 TeV
or heavier. In this case, the interaction through the magnetic 
moment can be very important and, for light nuclei, can even dominate 
$Z$-boson exchange. It is interesting to note that the electric 
dipole moment of a WIMP is also 
very constrained. The limit of $5
\, 10^{-21}$ e cm is stronger than any other 
EDM constraint, apart from EDMs of electrons, neutrons and heavy atoms. 
The method we used to obtain this limit is 
in the spirit of the original paper by Purcell and Ramsey \cite{PR},
where the first bound on the neutron EDM was inferred from neutron-nucleus
scattering. 

A model independent approach to the physics of WIMPs is 
partly motivated by recently discovered problems within the cold dark matter
model. While large-scale structure can be described reasonably well, 
the numerical simulations of galactic substructures appear to be in 
conflict with observational data \cite{problem}. A recently proposed cure,
invoking self-interacting dark matter \cite{selfint}, 
is incompatible with the very restrictive neutralino model. This fact
gives further impetus 
to our approach to study  WIMPs in a manner as model independent as 
possible. The limits on the electro-magnetic form factors obtained in 
this work show that the amount of self-interaction, which could be 
induced by these form factors, is not sufficient to generate a cross 
section of  WIMP-WIMP 
scattering at the level of $10^{-25}$ cm$^2$ as required by the
self-interacting WIMP scenario. 

The authors would like to thank Keith Olive, Arkady Vainshtein and Viktor Zacek 
for usefull discussions. This work was supported in part 
by the Department of Energy under Grant No.\ DE-FG-02-94-ER-40823.

\bibliographystyle{unsrt}

\end{document}